\documentclass[sigconf,nonacm]{acmart}
\AtBeginDocument{%
  \providecommand\BibTeX{{%
    \normalfont B\kern-0.5em{\scshape i\kern-0.25em b}\kern-0.8em\TeX}}}

\newcommand{\streamDaQ}{\textsc{Stream DaQ}}
\newcommand{\nameExplanation}{\underline{\textsc{S}tream}ing \underline{\textsc{D}a}ta \underline{\textsc{Q}}uality}

\usepackage{todonotes}
\setuptodonotes{color=blue!30}

\usepackage{listings}
\usepackage{xcolor}
\usepackage{enumitem} 
\usepackage{amsmath} 

\theoremstyle{definition}
\newtheorem{definition}{Definition}[section]

\lstdefinestyle{pythonstyle}{
    language=Python,
    captionpos=b,
    basicstyle=\ttfamily\small,
    keywordstyle=\color{blue},
    stringstyle=\color{violet},
    commentstyle=\color{olive},
    numbers=left,
    numberstyle=\tiny\color{gray},
    numbersep=5pt,
    frame=,
    breaklines=true,
    showstringspaces=false
}

\newcommand\numberthis{\addtocounter{equation}{1}\tag{\theequation}} 
\newcommand\gapBetweenPars{1 mm}

\newcommand{\tablesStretch}{\def\arraystretch{1.25}}
\newcommand{\rotateHeaders}{0}
\usepackage{pifont}  
\newcommand{\cmark}{\ding{51}}  
\newcommand{\xmark}{\ding{55}}  
\usepackage{multirow}
\usepackage{subfiles} 

\begin{document}

\title{\streamDaQ: Stream-First Data Quality Monitoring}

\author{Vasileios Papastergios}
\affiliation{%
  \institution{Aristotle University  of Thessaloniki}
  \city{Thessaloniki}
  \country{Greece}}
\email{papster@csd.auth.gr}

\author{Anastasios Gounaris}
\affiliation{%
  \institution{Aristotle University of Thessaloniki}
  \city{Thessaloniki}
  \country{Greece}
}
\email{gounaria@csd.auth.gr}

\renewcommand{\shortauthors}{V. Papastergios and A. Gounaris} %

\begin{abstract}
  Data quality is fundamental to modern data science workflows, where data continuously flows as unbounded streams feeding critical downstream tasks, from elementary analytics to advanced artificial intelligence models. Existing data quality approaches either focus exclusively on static data or treat streaming as an extension of batch processing, lacking the temporal granularity and contextual awareness required for true streaming applications. In this paper, we present a novel data quality monitoring model specifically designed for unbounded data streams. Our model introduces stream-first concepts, such as configurable windowing mechanisms, dynamic constraint adaptation, and continuous assessment that produces quality meta-streams for real-time pipeline awareness. To demonstrate practical applicability, we developed \streamDaQ, an open-source Python framework that implements our theoretical model. \streamDaQ~ unifies and adapts over 30 quality checks fragmented across existing static tools into a comprehensive streaming suite, enabling practitioners to define sophisticated, context-aware quality constraints through compositional expressiveness. Our evaluation demonstrates that the model's implementation significantly outperforms a production-grade alternative in both execution time and throughput while offering richer functionality via native streaming capabilities compared to other choices. Through its Python-native design, \streamDaQ~ seamlessly integrates with modern data science workflows, making continuous quality monitoring accessible to the broader data science community.
\end{abstract}


\maketitle

\section{Introduction}

\begin{figure}[tb!]
  \centering
  \includegraphics[width=0.8\linewidth]{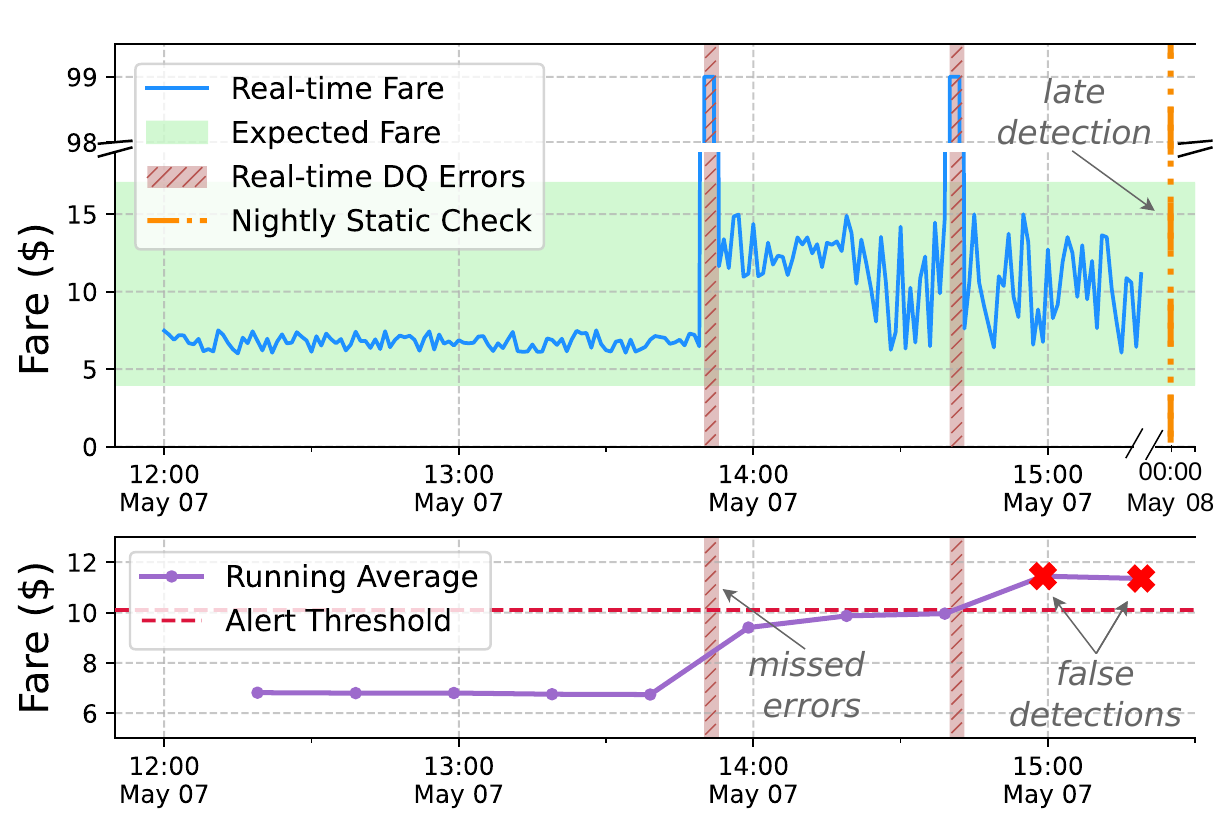}
  \caption{Real-time errors in NYC taxi data~\cite{nyc_taxi_dataset} can be detected late, falsely, or completely missed when applying static (top) or incremental (bottom) data quality checks.}
  \label{fig:late_detection_teaser}
\end{figure}

\begin{figure*}[tb!]
  \centering
  \includegraphics[width=\linewidth]{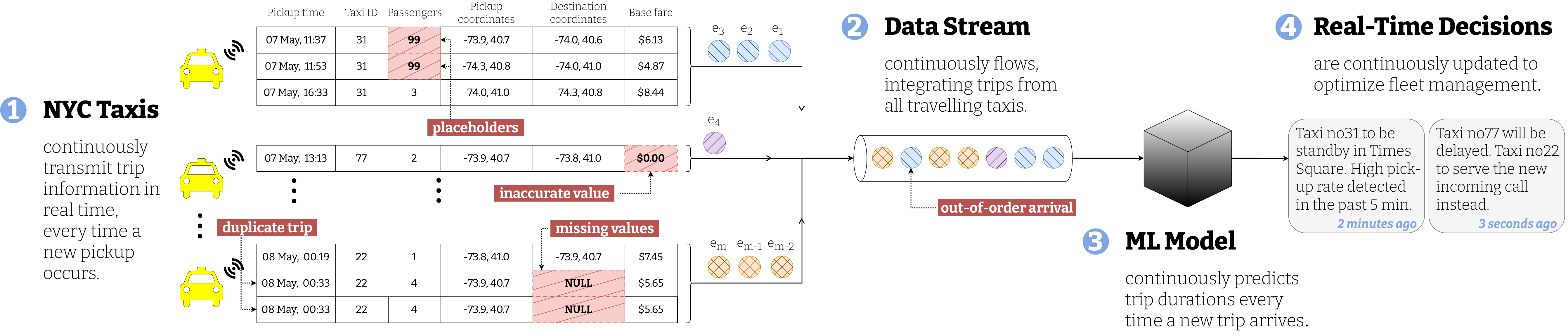}
  \caption{A real-world motivating example from the NYC taxi dataset~\cite{nyc_taxi_dataset} with generic and stream-specific data quality errors.}
  \label{fig:nyc_taxi_dq_example}
\end{figure*}

In an era of rapid machine learning (ML) advancements, data quality (DQ) is fundamental for data-driven operations~\cite{Agrawal2008, Abadi2016, cambridge_report_2025}. From business analytics to sophisticated artificial intelligence models, the quality of data consumed by these applications significantly impacts the reliability of their outcomes~\cite{mohammed_2025,Redman1998_impact_poor_dq_typical_enterprise, rein_edbt_2023,breck_data_validation_machine_learning}. As organizations increasingly shift towards real-time decision-making~\cite{Tien2017IoTrealTimeDecisionMaking, Koot2021IoTDecisionMaking}, the need for on-the-fly DQ monitoring becomes critical. High-volume, unbounded data streams from IoT devices~\cite{Cai2017, Gubbi2013IoT}, news feeds~\cite{liu2015newsFeed}, and financial transactions~\cite{Rajeshwari2016realTimeFraudDetection} are constantly processed for added value~\cite{stream_mining_2005}. Due to their volume and variety, data streams come with a wide range of data quality issues, including not only generic data errors~\cite{karlas_2024} present in all data regimes, but also temporally dependent ones~\cite{icewafl}. If left undetected, such issues can rapidly propagate through data pipelines, causing cascading errors and misguided business decisions~\cite{Wang2018, Rana2021_operational_inefficiencies, Loshin2011_business_impacts_of_dq}. Unfortunately, existing quality assessment approaches can not only leave such errors undetected, but also provide misleading detections. Figures ~\ref{fig:late_detection_teaser} and ~\ref{fig:nyc_taxi_dq_example} illustrate such a real-world example, discussed in detail in the rest of the section.

\vspace{\gapBetweenPars}\noindent\textbf{Limitations of Previous Approaches.} 
Although well-established, current quality assessment methodologies~\cite{Batini2009, Wang_1998, Batini2016} and implementations~\cite{Coleman_2013, survey_ehringer} address a different problem. They adopt either \textbf{static} or \textbf{incremental}~\cite{Schelter2019} data models, but not true \textbf{streaming} ones. The \emph{static} model performs one-off quality assessments on stored, unchanged data. The \emph{incremental} model treats streaming as an extension of batch processing, considering batch updates to evolving datasets that eventually become complete. Both models lack the necessary timeliness, temporal granularity and contextual awareness required for streaming applications, where the notion of a \emph{complete dataset} is typically not applicable.

\vspace{\gapBetweenPars}\noindent\textbf{Motivating Example.} 
~\autoref{fig:nyc_taxi_dq_example} illustrates a real-world example where existing data quality assessment approaches miss or misdetect real-time quality challenges posed by unbounded data streams. Consider the NYC taxi trip dataset~\cite{nyc_taxi_dataset} from an official, prized Kaggle competition. In this dataset, records of taxi rides, including pickup times and locations, destination coordinates, base fare amounts, and driver-reported passenger counts, arrive as a continuous stream. The competition called for an ML model that predicts trip durations in real time, informing fleet management decisions.

\vspace{\gapBetweenPars}\noindent\textbf{What Can Go Wrong? (\autoref{fig:nyc_taxi_dq_example}).}
Diverse, dynamic quality issues can arise in real-time streams~\cite{icewafl}. Sensor malfunctions might cause missing destination coordinates for temporally adjacent records. This renders the model's predictions for that period not only unreliable but also potentially misleading. Alternatively, drivers might fail to report the passenger count, leading to placeholder values (e.g., '99') for a series of rides. Away from the distribution the model was trained on, such values can directly impact duration predictions and the soundness of subsequent decisions. Moreover, network glitches can cause taxi ride records (i) to be duplicated within the stream, (ii) to be entirely lost for a period, or (iii) to arrive out of order. Contextual inconsistencies are also common: while typical NYC base fares average \$5.00-\$10.00 depending on traffic, errors might cause bursts of \$0.00 or \$100.00 fares. These real-time issues are not just isolated anomalies; they represent common data quality challenges that, if not timely detected, severely compromise the integrity of real-time insights and downstream applications.

\vspace{\gapBetweenPars}\noindent\textbf{Why Existing Approaches Would Fail? (\autoref{fig:late_detection_teaser}).}
A \textbf{static} data quality assessment system (top) running a batch job, perhaps nightly, would only detect these issues hours later. However, discovering missing destination coordinates or duplicate trips in yesterday's data is of little use for today's real-time fleet operations. By then, dispatchers might have sent taxis based on inaccurate estimates for an entire shift. Also, the opportunity for immediate correction is lost. An \textbf{incremental} data approach (bottom), although more flexible, also suffers fundamental limitations in a true streaming context. Consider the base fare scenario: if a temporary error introduces a burst of placeholder \$99.00 fares, an incremental model computing a running average since the start of the stream might miss it or only detect a gradual, delayed rise in the average fare. This detection would often occur after the period of actual low-quality data has ended, raising an alert when the quality is improved. Similarly, if a sensor stops reporting destination coordinates, an incremental model focused on total aggregates might only notice a slight, if any, growing percentage of NULL values, due to the masking effect of outdated, accumulated data. This aggregation, albeit suitable for slowly changing, finite datasets, contradicts a fundamental principle in data streams: \emph{data changes all the time}~(\citet{janus_project_2022}), so the latest information often matters the most.

\vspace{\gapBetweenPars}\noindent\textbf{Contributions.} 
To bridge this gap, we introduce (1) a novel data quality monitoring model for unbounded data streams, and (2) \streamDaQ~ (\nameExplanation), an open-source Python framework implementing our model. With a comprehensive suite of more than 30 temporally fine-grained quality checks, \streamDaQ~ can timely identify problems such as ``20\% missing destinations in the past 5 minutes'' or ``unusually high fare for specific taxi's last 5 rides within hourly context''. Our contributions are the following:
\begin{itemize}
  \item We formalize a novel data quality monitoring model based on configurable windowing mechanisms for timely, context-aware quality assessment of data streams (\autoref{sec:model});
  
  \item We present \streamDaQ, a scalable, open-source Python framework instantiating our model. \streamDaQ~ integrates and adapts over 30 fragmented quality checks from popular static tools into a rich, extensible streaming suite (\autoref{sec:implementation});

  \item We provide an extensive experimental analysis, comparing \streamDaQ~ against production-grade systems. Our evaluation demonstrates Stream DaQ's significant performance advantages and practical utility across various real-world stream quality tasks; e.g., for small windows improvements are up to 13.8x (\autoref{sec:experiments});

  \item  We demonstrate \streamDaQ's seamless, plug-and-play integration into existing data science workflows, highlighting its accessibility for data practitioners (\autoref{sec:how_to_use}); and

  \item We investigate rapidly emerging Python-based stream processing frameworks motivated by encountered implementation challenges, raising awareness towards their strengths and limitations from a practical standpoint (\autoref{sec:why_streaming_python}).
\end{itemize}

We start with preliminary concepts~(\autoref{sec:preliminaries}) and conclude with a review on related work (\autoref{sec:background_related_work}) and an outlook to future research directions~(\autoref{sec:conclusion}). Our source code has been made publicly available under an open-source license at \url{https://github.com/Bilpapster/Stream-DaQ}.

\section{Preliminaries}
\label{sec:preliminaries}
In this section, we formally present preliminary information relevant to data streams and data quality, reviewing theoretical definitions (\autoref{subsec:definitions}) and background concepts (\autoref{subsec:concepts}).

\subsection{Definitions}
\label{subsec:definitions}

\begin{definition}[Data Stream Element]
  A data stream element is a tuple $e = \langle t, A_1, A_2, \dots, A_k \rangle$ containing a timestamp $t$ and $k\in\mathbb{N}$ attributes $A$,
  where $t$ corresponds to the element's creation time (\emph{event time}). Each attribute has values from a domain: $A_i\in dom(A_i),\;i\in[1, k]$. For example, in the NYC taxi dataset of~\autoref{fig:nyc_taxi_dq_example}, a data stream element is a pickup event transmitted by a taxi $e_{pickup} = \langle \text{May-07 11:37},\; 31,\; 3, \dots,\; \$6.13 \rangle$, where $t$ is the pickup time, $A_1$ the taxi id, $A_2$ the passenger count, and so on.
\end{definition}

\begin{definition}[Data Stream]
  A data stream is a sequence $D = e_1, e_2, \dots, e_N$ of $N\in\mathbb{N}$ data stream elements. For finite streams, $N$ is known and $|D|=N$. For unbounded streams, $N$ is unknown and the sequence comprises potentially infinite elements arriving over time. We focus on unbounded data streams to cover the complete spectrum of real-world streaming applications, though our model applies to finite streams without modification.
\end{definition}

\begin{definition}[Stream Window] 
  \label{def:window}
  A stream window $W$ is a finite partition of the data stream containing all elements whose timestamps fall within the range defined by a $start$ and $end$ timestamps:
  \begin{equation}
    \label{eq:window}
    W_{[start,\;end]} = \{ e_i \; \forall \; i\in[1, n]: e_i.t \in [start, end] \}
  \end{equation}
  where $e_i.t$ refers to the timestamp of the $i$\textsuperscript{th} data stream element. Windows enable meaningful, space-bounded computations over unbounded streams~\cite{li_2005_stream_windows}. For example, in the NYC taxi dataset, a window $W_{[\text{07 May 11:30,\; 07 May 13:30}]} = \{e_1, e_2, e_4\}$ includes elements whose pickup timestamps fall within the defined 2-hour range.
\end{definition}

\begin{definition}[Static Data Quality Check]
  \label{def:static_checks}
  Given a static dataset $\mathcal{R}_{static}=\{r_i\}_{i=1}^N$ of $N\in\mathbb{N}$ records with $k$ attributes each, and a constraint $\mathcal{C}$, a static data quality check is a function that maps $\mathcal{R}_{static}$ and $\mathcal{C}$ to a value $v$:
  \begin{equation}
    \label{eq:dq_check_static}
    Check_{static}(\mathcal{R}_{static}, \mathcal{C}) = v, \;v\in\{True, False\} \cup \mathbb{N}
  \end{equation}
  where $v$ can be boolean (constraint satisfaction) or numerical (count of records meeting the constraint). Comparing numerical values against thresholds produces boolean assessments. For example, assuming a static dataset $\mathcal{R}_{static}=\{e_i\}_{i=1}^N$ from the NYC taxi data:
  \begin{displaymath}
    Check_{static}(\mathcal{R}_{static}, \;e_i.passengers\in[1, 4] \forall i\in[1, N]) = False
  \end{displaymath}
  validates whether passenger counts are plausible for every ride. Note that static records $r_i$ may lack timestamps, while stream elements $e_i$ require them.
\end{definition}

\begin{definition}[Incremental Data Quality Check]
\label{def:incremental_checks}
An incremental data quality check operates on an \emph{evolving} dataset $\mathcal{R}_{evolv}$ formed by the union of $b\in\mathbb{N^+}$ non-overlapping batches:
\begin{equation}
    \label{eq:incremental_dataset}
    \mathcal{R}_{evolv}^{(b)} = \bigcup_{i=1}^b B_i = \bigcup_{i=1}^b\Delta\mathcal{R}_{evolv}^{(i)}
\end{equation}
where $\mathcal{R}_{evolv}^{(i)}$ is the dataset snapshot after appending $i$ batches to the initial empty snapshot $\mathcal{R}_{evolv}^{(0)} = \emptyset$, and $\Delta\mathcal{R}_{evolv}^{(i)}$ represents newly appended records in $B_i$. Batches are disjoint: $B_i \cap B_j = \emptyset \;\forall\; j\neq i\in[1, b]$. The final snapshot is the \emph{complete dataset} $\mathcal{R}_{evolv}^{(b)} \equiv\mathcal{R}_{static}$.

Introduced by \citet{Schelter2019} for append-only updates, incremental checks produce results equivalent to static checks while processing only the new batch $B_i$ each time, keeping some values as \emph{state} $S$ to avoid unnecessary computations for previous batches:
  \begin{align*}
    Check_{increm}(\mathcal{R}_{evolv}^{(b)}, \mathcal{C})  &= \{Check_{static'}(B_i, S^{(i-1)}, C)\}_{i=1}^b \\
                                                            &= \{\langle v_i, S^{(i)} \rangle\}_{i=1}^b = v_b \numberthis
  \end{align*}
where $S^{(i)}$ contains computed values after $i$ batches and $Check_{static'}$ is a transformed static check that accepts the previous state $S^{(i-1)}$ and produces a tuple $\langle v_i, S^{(i)} \rangle$ with result and updated state. The final result $v_b$ equals the static check result on $\mathcal{R}_{static}$.

For example, in ~\autoref{fig:late_detection_teaser} (bottom), a running fare average check on evolving taxi data $\mathcal{R}_{evolv}=\bigcup_{i=1}^bB_i$, where $B_i \subset \{e_i\}_{i=1}^N$ is:
  \begin{displaymath}
    Check_{increm}(\mathcal{R}_{evolv}, \;\overline{e_i.fare} \leq 10) = \{True, \dots, False\} = False
  \end{displaymath}
  where state $S^{(i)}$ keeps (running) sum and count of fares in $\mathcal{R}_{evolv}^{(i)}$.
\end{definition}

\subsection{Background Concepts}
\label{subsec:concepts}

We discuss here fundamental stream processing (\autoref{subsubsec:stream_processing}) and data quality management (\autoref{subsubsec:data_quality_management}) concepts, focusing on \emph{monitoring}, the foundation of our streaming data quality model.

\subsubsection{Stream Processing Fundamentals}
\label{subsubsec:stream_processing}
Stream processing systems handle potentially infinite sequences of data elements arriving continuously, typically under strict latency requirements and resource constraints~\cite{Maier_2004_semantics_data_streams, marcos_2018}. Unlike batch processing on finite datasets, stream processing requires real-time data handling. \emph{Windowing} (see \autoref{def:window}) enables bounded computations over unbounded streams~\cite{li_2005_stream_windows}.

Windows are broadly categorized as \emph{time-}, \emph{count-}, or \emph{session-based}~\cite{akidau_2015_datastreams_google, Carbone2015_flink}. \emph{Time-based} windows group elements by timestamps using overlapping (\emph{sliding}) or disjoint (\emph{tumbling}) intervals~\cite{li_2005_stream_windows}. \emph{Count-based} windows group fixed numbers of elements. \emph{Session-based} windows group elements based on activity periods separated by inactivity gaps; suitable for analyzing event-driven data streams~\cite{akidau_2015_datastreams_google}. Count- and session-based windows can be transformed to~\autoref{eq:window} using timestamps of their earliest and latest elements.

Modern stream processing frameworks, thoroughly discussed in~\autoref{subsec:traditional_stream_processing}, implement windowing differently. Apache Flink provides native event-time windows with watermark mechanisms~\cite{Carbone2015_flink}. Spark Streaming uses micro-batch architecture~\cite{Zaharia2013}. Apache Storm offers time and count windows through its core API~\cite{Toshniwal2014}. Kafka Streams maintains window state through local stores~\cite{kafka_streams}. In \streamDaQ, we leverage these windowing concepts as the foundation for temporally fine-grained stream quality monitoring, enabling context-aware assessment over configurable time horizons.

\subsubsection{Data Quality Management}
\label{subsubsec:data_quality_management}
Data quality management encompasses \emph{measurement}, \emph{assessment}, \emph{monitoring}, and \emph{improvement}~\cite{Batini2016}. Drawing inspiration from how measurement and assessment applies to static data, our model meticulously adapts these concepts to fit streaming applications. We thereafter build on top of these adaptations in \streamDaQ's key novelty: continuous monitoring of unbounded data streams for timely, on-the-fly error detection.

\vspace{\gapBetweenPars}\noindent\textbf{Data quality \emph{measurement}}
involves computing specific metrics that quantify data quality aspects~\cite{survey_ehringer} such as completeness, accuracy, or consistency, also knwon as \emph{dimensions} or \emph{characteristics}~\cite{mohammed_glossary_2024, iso25012}. For example, counting NULL values contributes to quantifying completeness~\cite{Batini2016}. These metrics, often derived from data profiling~\cite{Abedjan2015}, provide raw numerical values, albeit without inherent indication whether data meets desired quality standards.

\vspace{\gapBetweenPars}\noindent\textbf{Data quality \emph{assessment}}
evaluates measurements against predefined rules or thresholds, deriving from general knowledge, reference datasets, or standards~\cite{Coleman_2013}. Through \emph{data quality checks} (see Def. \ref{def:static_checks} and \ref{def:incremental_checks}), the goal of assessment is to draw conclusions about whether the data at hand is fit for use by humans or downstream tasks~\cite{WangStrong1996}. For example, while measurement might reveal that 95\% of values are non-missing, assessment involves deciding whether this suffices for the intended use. Assessment is inherently context-dependent, as different domains and applications may have varying quality requirements~\cite{Strong1997, mohamed2024Multifaceted, fadlallah_data_quality_context_2023, serra_data_quality_context_2024}. Overall, assessment transforms raw measurements into actionable quality insights.

\vspace{\gapBetweenPars}\noindent\textbf{Data quality \emph{monitoring}} 
in static and incremental models extends assessment by periodically evaluating quality rules over time~\cite{survey_ehringer, Ehrlinger2023}; for example, static checks on the whole dataset repeated nightly. Data streams, however, have received significantly less attention, despite the fact that, in real-time pipelines, issues must be detected and addressed promptly to prevent downstream impacts~\cite{Schelter2018}. In streaming contexts, one-off assessments are of limited utility due to the dynamic nature of data and ephemeral yet impactful quality issues~\cite{icewafl}. Traditional monitoring approaches relying on batch updates are inherently insufficient for unbounded streams where a dataset may never be complete. Given the high volatility and potential concept drifts in data streams~\cite{concept_drift_2014}, recent data often carries greater relevance than historical observations~\cite{janus_project_2022}. Our model (\autoref{sec:model}) addresses this gap through continuous quality assessment over configurable temporal windows that can also access contextual information in user-driven granularities.

\vspace{\gapBetweenPars}\noindent\textbf{Data quality \emph{improvement}}
(also found as \emph{data cleaning}, \emph{debugging}, or \emph{repair}) represents the ultimate goal of quality management. It encompasses actions taken to enhance data quality based on insights from measurement, assessment, and monitoring. Modern tools ~\cite{activeClean2016, katara2015, nadeef2013} and techniques~\cite{wang_imputation_2024, firmani_entity_resolution_2016} implement improvement through operations like missing data imputation or entity reconciliation. Our proposed model, although not directly touching data quality improvement, aims to inform and advance it through \emph{timely} and \emph{accurate} detection of quality issues reliably over time.

\section{A Data Quality Model Designed For Streams}
\label{sec:model}

In this section, we formally present our novel data quality monitoring model for unbounded data streams. We begin by defining the problem (\autoref{subsec:problem_definition}) and introduce the fundamental building blocks that adapt traditional measurement and assessment concepts to streaming contexts (\autoref{subsec:fundamental_components}). We then elaborate on the categories of streaming quality checks that instantiate our model (\autoref{subsec:check_categories}), highlighting their expressiveness through composition (\autoref{subsec:check_composition}). Throughout, we revisit the motivating example of ~\autoref{fig:nyc_taxi_dq_example} to help readers better comprehend our model.

\subsection{Problem Definition}
\label{subsec:problem_definition}

We address the problem of continuous, real-time data quality monitoring over unbounded data streams. Given an unbounded data stream $D = e_1, e_2, \dots$ and a set of quality constraints $\mathcal{C}$, our goal is to continuously assess data quality as new elements arrive, providing timely, temporally fine-grained detection of both generic and temporally dependent data errors. No assumptions are made that the dataset is or will become complete. The monitoring process continues indefinitely as long as data arrives. When quality issues are detected, alerts are raised and users can define remedial actions (e.g., routing erroneous elements to separate streams), but monitoring never terminates. This way, we address the dynamic nature of streams where quality issues may be ephemeral yet impactful, requiring immediate detection and response~\cite{icewafl}.

\subsection{Fundamental Components}
\label{subsec:fundamental_components}

We formalize here how traditional data quality concepts adapt to streaming contexts by introducing streaming-specific definitions for measurement, assessment, and monitoring.

\begin{definition}[Stream Quality Measurement]
\label{def:stream_measurement}
Given a stream window $W_{[start,\;end]}$ (\autoref{def:window}) and a quality metric $m$, stream quality measurement computes a numerical value $v_m$:
\begin{equation}
\label{eq:stream_measurement}
Measure_{stream}(W_{[start,\;end]}, m) = v_m \in \mathbb{R}
\end{equation}
where $v_m$ represents the measured value for metric $m$ over the window. For example, measuring the fraction of non-NULL passenger counts (completeness) in a 5-minute window can be written as
    $Measure_{stream}(W_{[11:30,\;11:35]}, completeness_{passengers}) = 0.85$.
\end{definition}

\begin{definition}[Stream Quality Assessment]
\label{def:stream_assessment}
Stream quality assessment first computes measurements over a window, then evaluates them against quality constraints:
\begin{equation}
\label{eq:stream_assessment}
Assess_{stream}(W_{[start,\;end]}, \mathcal{C}_{stream}) = \langle v_m, v_a \rangle
\end{equation}
where $v_m = Measure_{stream}(W_{[start,\;end]}, m)$ for the relevant metric $m$ specified in constraint $\mathcal{C}_{stream}$, and $v_a \in \{True, False\} \cup \mathbb{R}$ represents the assessment result obtained by comparing $v_m$ against the constraint threshold. If assessment alone is considered, $v_a$ is sufficient. For instance, assessing if completeness exceeds 90\%:
$Assess_{stream}(W_{[11:35,\;11:40]}, completeness_{passengers} \geq 0.9) = \langle 0.75, False \rangle$,
where 0.75 is the measured completeness and $False$ indicates the assessment outcome (constraint violation).
\end{definition}

\begin{definition}[Stream Quality Monitoring]
    \label{def:stream_quality_monitoring}
    Building on these primitives, we define streaming data quality monitoring as the continuous application of assessments over a sequence of windows:
    \begin{align*}
    Check_{stream}(D, \mathcal{C}_{stream}) &= \{Assess_{stream}(W_i, \mathcal{C}_{stream}) \; \forall \; W_i \subseteq D\} \\
                                     &= \{\langle v_{m_1}, v_{a_1} \rangle, \langle v_{m_2}, v_{a_2} \rangle, \dots, \langle v_{m_l}, v_{a_l} \rangle\} \numberthis
    \end{align*}
    where $W_i$ represents the $i$\textsuperscript{th} window of stream $D$, each tuple $\langle v_{m_i}, v_{a_i} \rangle$ contains the measurement and assessment results for window $W_i$, and $l \in \mathbb{N}$ is the total number of windows processed. For unbounded streams, $l$ grows indefinitely.
\end{definition}
This formulation fundamentally differs from both static checks (\autoref{def:static_checks}), which operate in an one-off manner on unchanged datasets, and incremental ones (\autoref{def:incremental_checks}), which assume appendices towards dataset completeness. The following key differences hold:
    
\begin{itemize}[leftmargin=*]
    \item \textbf{Fine-grained temporal scope}: Checks operate on configurable temporal windows with a dual benefit: only the most relevant data is 
    monitored, with relevancy being user-driven, while erasing the inapplicable dependency on dataset completeness;
    \item \textbf{Dynamic constraints}: Dynamic constraint $\mathcal{C}_{stream}$ can adapt based on contextual information from previous windows at user-driven granularities (detailed in \autoref{subsec:check_categories}); and
    \item \textbf{Sequential results}: Output is a sequence of measurement-assessment tuples enabling continuous monitoring rather than single-point, one-off assessment. This introduces another novel concept of our model, the \emph{quality meta-stream}. 
\end{itemize}

\begin{definition}[Quality Meta-Stream]
    \label{def:meta_stream}
    Given a streaming quality check $Check_{stream}(D, \mathcal{C}_{stream})$ that produces a sequence of measurement-assessment tuples, we define the corresponding quality meta-stream $\mathcal{M}$ as:
    \begin{equation}
    \label{eq:meta_stream}
    \mathcal{M}(D, \mathcal{C}_{stream}) = \langle t_{start_i}, t_{end_i}, v_{m_i}, v_{a_i} \rangle_{i=1}^l
    \end{equation}
    where $t_{start_i}$ and $t_{end_i}$ represent the start and end timestamps respectively of window $W_i$, $v_{m_i}$ is the measurement result, $v_{a_i}$ is the assessment result, and $l$ is the number of processed windows. The meta-stream $\mathcal{M}$ transforms quality monitoring results into a structured stream that can be processed by downstream systems and act as a quality-awareness signal for the whole pipeline. Inspired by the recent advancements in the area of learning from imperfect data (overviewed by ~\citet{karlas_2024}), we aspire that this meta-stream could advance quality awareness in online downstream applications, enabling them to not only operate reliably, but also continue their training, even in the presence of real-time data errors. 
    
    For example, monitoring average taxi fares over sliding 10-minute windows with 1-minute slides produces
    \begin{align*}
        \mathcal{M}(D_{taxi}, \overline{fare} \leq \$10) = \{ & \langle 11:35,\; 11:40,\; \$9.29,\; True\rangle, \\
                                                           & \langle 11:36,\;11:41,\;\$10.56,\; False\rangle,\; \dots\}
    \end{align*}
    This meta-stream can flow through the fleet management pipeline, potentially informing (i) the human operator that model decisions (or suggestions) can be suboptimal due to data errors, and (ii) the ML model to temporarily skip predictions or weight updates. Triggering on-the-fly remedial actions is detailed in ~\autoref{sec:implementation}.
\end{definition}

\subsection{Categories of Streaming Quality Checks}
\label{subsec:check_categories}
We next present further refinements of our streaming quality check template (\autoref{def:stream_quality_monitoring}) to support diverse monitoring needs. Acknowledging that streaming applications vary significantly in their quality requirements, we provide an expressive suite of check categories that enable users to define sophisticated, context-aware constraints. These categories are not mutually exclusive and can be combined to create composite checks for complex quality scenarios.

\subsubsection{Tuple-at-a-Time Checks}
\label{subsubsec:tuple_checks}
Tuple-at-a-time checks evaluate each stream element independently within a window:
\begin{equation}
\label{eq:tuple_check}
Check_{tuple}(W, \mathcal{C}) = \{Assess_{stream}(\{e_i\}, \mathcal{C}) \; \forall \; e_i \in W\}
\end{equation}
where each element $e_i$ is assessed in isolation. By setting very short window lengths, these checks enable (near-) real-time error detection. For example, validating that pickup and destination coordinates differ for each taxi trip can be written as:
\begin{align*}
Check_{tuple} (W_{[11:30,11:35]},\; pickup\_coords \neq dest\_coords) = \\ 
= \left\{ \{True, True, False, \dots\},\; \dots \right\}
\end{align*}
where a $False$ result immediately identifies a problematic trip without requiring context from other elements. 

\subsubsection{Window Context Checks}
\label{subsubsec:window_checks}
Window context checks require all elements within a window to produce results:
\begin{equation}
\label{eq:window_check}
Check_{window}(W, \mathcal{C}) = Assess_{stream}(W, \mathcal{C})
\end{equation}
where constraint $\mathcal{C}$ operates on the entire window $W$ rather than individual elements. Examples include detecting sensor malfunctions through identical values:
\begin{displaymath}
Check_{window}(W_{[11:30,11:35]}, |unique(passenger\_count)| > 1) = True,
\end{displaymath}
or validating average fare thresholds:
\begin{displaymath}
Check_{window}(W_{[11:30,11:35]}, \overline{base\_fare} \leq \$10.00) = True
\end{displaymath}
Again, very small windows enable (near-) real time detection.

\subsubsection{Reference Data Checks}
\label{subsubsec:reference_checks}
Reference data checks incorporate external datasets as constraint baselines:
\begin{equation}
\label{eq:reference_check}
Check_{ref}(W, \mathcal{C}, Ref) = Assess_{stream}(W, \mathcal{C}(Ref))
\end{equation}
where $Ref$ represents reference data and $\mathcal{C}(Ref)$ is a constraint parameterized by $Ref$. For instance, validating current fares against historical hourly distributions:
\begin{displaymath}
Check_{ref}(W_{[15:30,15:35]}, |\overline{fare}_W - \overline{fare}_{Ref_{15h}}| \leq \$2.00, Ref_{hist}) = True
\end{displaymath}
where $Ref_{hist}$ contains reference fare distributions and $Ref_{15h}$ denotes the 3 PM reference distribution.

\subsubsection{Dynamically Adapted Context Checks}
\label{subsubsec:dynamic_checks}
These checks adapt constraints based on contextual information from extended, configurable time horizons to fit diverse streaming applications:
\begin{equation}
\label{eq:dynamic_check}
Check_{dynamic}(W, \mathcal{C}, H) = Assess_{stream}(W, \mathcal{C}(Context(H)))
\end{equation}
where $H$ defines a time horizon for context computation and $Context(H)$ extracts contextual state over that horizon. For example, detecting fare deviations using rolling hourly context can be written as:
\begin{displaymath}
Check_{dynamic}(W_{[11:30,11:35]}, |\overline{fare}_W - \mu_H| > 3\sigma_H, H_{1hour}) = False
\end{displaymath}
where $\mu_H$ and $\sigma_H$ represent mean and standard deviation computed over the past hour. This approach enables detection of temporally localized errors prevalent in streams~\cite{icewafl} while maintaining user-controlled temporal granularity; fundamentally different from incremental models that aggregate since stream inception.

\subsubsection{Keyed Stream Checks}
\label{subsubsec:keyed_checks}
Keyed checks partition streams by attribute values, applying windowing and quality constraints within each partition:
\begin{equation}
\label{eq:keyed_check}
Check_{keyed}(W, \mathcal{C}, key) = \{Check_{stream}(W_k, \mathcal{C}) \; \forall \; k \in keys(W)\}
\end{equation}
where $W_k = \{e_i \in W : e_i.key = k\}$ represents the window subset for key value $k$. For example, detecting consecutive missing destinations per taxi can be written as:
\begin{displaymath}
Check_{keyed}(W_{[11:00,11:30]}, consecutive\_nulls(dest) \geq 5, taxi\_id),
\end{displaymath}
which produces separate assessments for each taxi, enabling fine-grained, entity-specific quality monitoring.

\subsection{Expressiveness Through Composition}
\label{subsec:check_composition}

The true power of our stream quality monitoring model emerges through the composition of check categories. By combining different categories, users can define sophisticated quality constraints that address complex real-world monitoring scenarios. Revisiting our motivating example (\autoref{fig:nyc_taxi_dq_example}), we present here several indicative examples of such composite checks supported by our model.

\vspace{\gapBetweenPars}\noindent\textbf{Keyed + Dynamic Context.}
Combining keyed partitioning with dynamic context enables entity-specific monitoring with configurable temporal awareness. For example,
\begin{displaymath}
Check_{keyed+dynamic}(W, |\overline{fare}_{W_k} - \mu_{H_k}| > 2\sigma_{H_k}, taxi\_id, H_{2hours})
\end{displaymath}
detects per-taxi fare inaccuracies using each taxi's individual 2-hour fare history. Using a 2-hour context enables keeping only recent, relevant information, accounting for potential changes in the traffic\footnote{It is common that taxi base fares are surcharged based on the amount of time the vehicle is stopped or moving slowly (e.g., below 12 mph).}. The window duration can be significantly shorter, e.g., 5 minutes, allowing for on-the-fly detection when a specific taxi reports unusually high or low fares compared to its recent patterns. 

\vspace{\gapBetweenPars}\noindent\textbf{Tuple-at-a-time + Reference Data.}
Combining tuple-level validation with reference data enables error detection using established external standards. For example,
\begin{displaymath}
Check_{tuple+ref}(W, distance(pickup, dest) \in valid\_routes_{Ref})
\end{displaymath}
validates each trip's pickup-destination pair against reference information of valid NYC coordinates and routes $valid\_routes_{Ref}$, flagging in real time impossible or potentially erroneous routes.

\vspace{\gapBetweenPars}\noindent\textbf{Window + Keyed + Dynamic Context.}
Such a triple composition enables sophisticated contextual monitoring. For example,
\begin{align*}
&Check_{window+keyed+dynamic}(W, \\
&\qquad \frac{|null\_destinations_{W_k}|}{|W_k|} > 3 \times \frac{|null\_destinations_{H_k}|}{|H_k|}, \\
&\qquad taxi\_id, H_{1hour})
\end{align*}
detects when the rate of missing destinations for a specific taxi in the current window exceeds three times its historical hourly rate, indicating potential sensor or network connectivity degradation for that particular vehicle. Informing the driver could result in timely correction (e.g., by rebooting the GPS device).

\section{The \streamDaQ~ Framework}
\label{sec:implementation}

This section presents \streamDaQ, an open-source Python implementation of our stream quality monitoring model. Built on the Pathway streaming framework~\cite{pathway2023, pathway_github}, \streamDaQ~ enables seamless integration into existing data science workflows through an intuitive interface accessible to the broader data science community. We present \streamDaQ's architecture (\autoref{subsec:architecture}), its comprehensive check suite (\autoref{subsec:check_suite}), and demonstrate practical integration (\autoref{subsec:integration}). Implementation choices and middleware decisions are detailed in \autoref{sec:why_streaming_python}.

\subsection{An Architecture Designed for Streams}
\label{subsec:architecture}

\begin{figure*}[tb!]
  \centering
  \includegraphics[width=\linewidth]{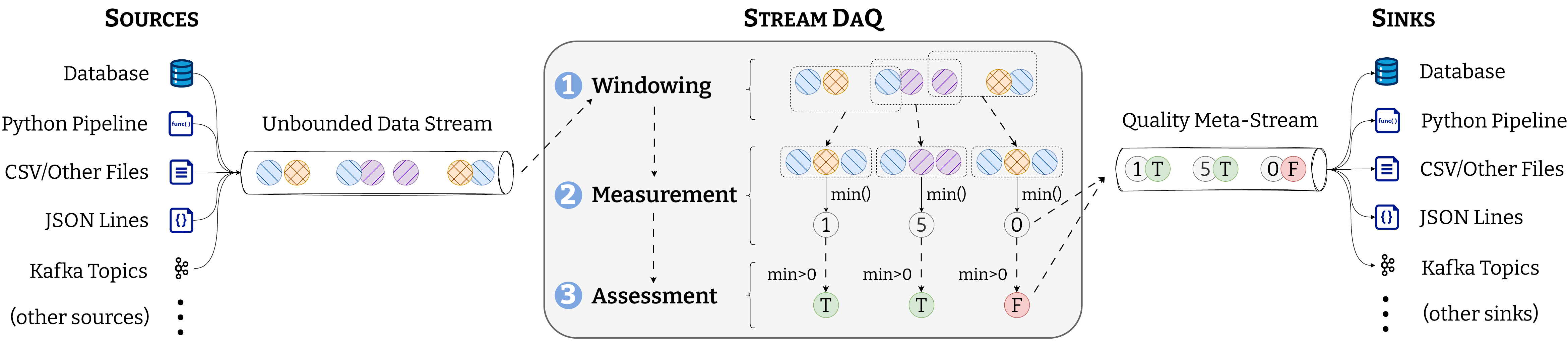}
  \caption{The \streamDaQ~ framework architecture.}
  \label{fig:architecture}
\end{figure*}

\streamDaQ's architecture, shown in~\autoref{fig:architecture}, directly implements the fundamental components of our model (\autoref{subsec:fundamental_components}). Configurable windowing mechanisms enable bounded computations over unbounded streams, a lightweight measurement engine computes quality metrics over these windows, and an assessment component continuously evaluates measurements against constraints, producing the quality meta-stream. This pipeline runs continuously as new elements arrive, realizing stream quality monitoring in practice.

\vspace{\gapBetweenPars}\noindent\textbf{Windowing.}
\streamDaQ~ implements windowing concepts (\autoref{def:window}) through abstractions that hide underlying complexity, while providing full configurability, such as straightforward access to window start and end timestamps. Users can define time-based (tumbling or sliding) and session-based windows with intuitive interfaces and sensible defaults. For example, practitioners can configure sliding 5-minute windows with 1-minute slides, or session windows grouping consecutive events separated by inactivity greater than 2 minutes. \streamDaQ~ also supports late-arrival handling through flexible cutoff mechanisms. For instance, practitioners might configure that stream elements arriving more than 3 minutes late are of little usefulness and should be discarded without being processed.

\vspace{\gapBetweenPars}\noindent\textbf{Stream Quality Measurement.}
The framework implements the concepts from~\autoref{def:stream_measurement} through a lightweight measurement engine that computes quality metrics over windowed data. \streamDaQ~ unifies and enriches more than 30 fragmented functionalities that appear across multiple widely used tools, after meticulously adjusting them from static to online contexts. These checks, overviewed in ~\autoref{tab:quality_checks}, are discussed in detail in~\autoref{subsec:check_suite}. In terms of implementation, \streamDaQ~ leverages both Pathway's built-in reducers\footnote{Adopting Pathway's naming conventions, \emph{reducers} refer to programmatic functions that operate on the data stream elements within a window.} (in Rust) and carefully designed custom reducers (in Python) to synthesize a hybrid, lightweight measurement engine. 

\vspace{\gapBetweenPars}\noindent\textbf{Continuous Assessment and Monitoring.}
\streamDaQ~ materializes stream quality assessment (\autoref{def:stream_assessment}) and monitoring (\autoref{def:stream_quality_monitoring}) through evaluation of flexible, user-driven constraints. The framework continuously applies assessment functions over measurement results, producing the meta-stream $\mathcal{M}(D, \mathcal{C}_{stream})$ of quality insights. This meta-stream can serve as a real-time quality awareness signal for the whole pipeline and is readily accessible through a few lines of Python code. When quality issues are detected, alerts are raised and users can define remedial actions such as routing problematic elements to separate streams for further investigation.

\begin{table*}[tb]
  \tablesStretch
  \caption{Built-in Quality Check Suite in \streamDaQ.}
  \label{tab:quality_checks}
  \resizebox{\linewidth}{!}{%
    \begin{tabular}{llll}
      \toprule
      \textbf{Category} & \textbf{Stream Quality Check}     & \textbf{Data Science Use Case Example}                                & \textbf{Configuration Options}                  \\
      \midrule
      \multirow{6}{*}{\rotatebox{60}{Tuple-at-a-time}} 
                     & valid range                       & Validate sensor readings are within acceptable boundaries           & inclusive/exclusive endpoints     \\
                     & accepted value sets               & Ensure categorical features match expected model classes                & proper/improper subset            \\
                     & pattern matching                  & Verify text inputs conform to expected format for NLP pipelines        & predefined/custom regex patterns \\
                     & row-wise conformance              & Validate feature vector completeness before ML inference               & predefined/custom SQL expressions \\
                     & value ordering                    & Validate strictly increasing transaction IDs or event sequences         & increasing/decreasing, strict or non     \\
                     & cross-interval validation         &Detect resource allocation conflicts in task scheduling systems     & gaps allowed/disallowed/required \\
      \midrule
      \multirow{17}{*}{\rotatebox{60}{Window Context}} 
                     & stream freshness                  & Ensure recommendation systems use recent user interaction data          & system/custom reference time      \\
                     & matching between streams          & Verify feature-label alignment in online learning pipelines            & normal/windowed join validation   \\
                     & dead stream detection             & Alert when real-time model inference pipeline stalls                   & configurable restart triggers    \\
                     & frozen stream detection           & Identify sensor malfunctions in IoT-based predictive models            & return-to-normal notifications    \\
                     & out-of-order arrival detection    & Maintain sequence integrity for RNN/LSTM model inputs                  & on event time/custom field            \\
                     & element length statistics         & Monitor text input lengths for transformer model compatibility          & length/statistical measures      \\
                     & distribution analysis             & Detect data drift in production ML model inputs                        & quartiles/custom percentiles     \\
                     & window statistics                 & Filter statistical outliers from real-time analytics                  & max/min/mean/std/z-scores        \\
                     & volume monitoring                 & Track batch size variations in distributed model training              & absolute/relative thresholds     \\
                     & distinct element counting         & Monitor user interactions within sessions in a recommend. system               & exact/approximate algorithms     \\
                     & uniqueness validation             & Detect duplicate samples in streaming model training data              & unique count/ratio computation   \\
                     & heavy hitters identification      & Identify dominant classes in imbalanced streaming datasets             & exact/approximate counting       \\
                     & missing stream elements            & Track feature completeness for automated feature engineering           & custom definition of missing \\
                     & placeholder consistency           & Ensure uniform null handling across multi-stage ML pipelines          & custom definition of placeholder   \\
                     & correlation analysis              & Monitor feature relationships in real-time feature stores             & Pearson/Spearman/mutual info     \\
                     & schema validation                 & Maintain feature consistency between training and inference stages     & presence/absence order of columns \\
                     & data type validation              & Prevent type mismatches in automated ML pipeline deployments          & formats (e.g., \texttt{dateutil})   \\
      \bottomrule
    \end{tabular}
  }
\end{table*}

\subsection{A Check Suite Designed for Comprehensiveness}
\label{subsec:check_suite}

At the heart of \streamDaQ~ lies an extensive suite of built-in quality checks that implement the measurement concepts from our model (\autoref{def:stream_measurement}). Designed to address the inherently context-dependent nature of data quality~\cite{Strong1997, fadlallah_data_quality_context_2023}, these checks represent a unified collection of quality assessments adapted from seven widely-used static data quality tools based on~\cite{DQarxiv}. The tools are: Apache Griffin~\cite{griffin_github}, dbt Core~\cite{dbt_github}, Deequ~\cite{deequ_github}, Evidently~\cite{evidently_github}, Great Expectations~\cite{gx_github}, MobyDQ~\cite{moby_github}, and Soda Core~\cite{soda_github}. All these checks, previously fragmented across static tools, have been meticulously collected and adapted to streaming contexts, enabling practitioners to leverage familiar quality concepts in real-time scenarios.

Data quality requirements can vary significantly across different scenarios. What constitutes \emph{good quality} for sensor data in industrial monitoring (e.g., sensor readings fall inside acceptable ranges) might differ substantially from quality requirements for financial transaction streams (e.g., customer ids match existing accounts). Even within the same domain, quality definitions can vary based on the specific task, system constraints, or end-user requirements. For example, in industrial monitoring, early detection of a frozen stream and monitoring readings distribution might be of heightened importance for predictive maintenance tasks~\cite{DIEZOLIVAN201992}, while ensuring low rates of missing information might be prioritized for real-time dashboards informing tactical business decisions. Acknowledging this context dependency, \streamDaQ~ provides data practitioners with a rich suite of quality checks they can selectively apply to their specific use cases. ~\autoref{tab:quality_checks} summarizes \streamDaQ's check suite.

The suite encompasses both tuple-at-a-time checks (operating on individual stream elements) and window context checks (requiring all elements within a window), as formalized in \autoref{subsec:check_categories}. Importantly, any check can be enhanced with reference data, dynamic context adaptation, or keyed partitioning, unlocking the power of compositional expressiveness demonstrated in \autoref{subsec:check_composition}. Beyond these built-in capabilities, \streamDaQ~ is designed for extensibility. Practitioners can easily implement custom checks by defining Python functions that accept windowed data and produce measurement results. Windowing, continuous assessment, and meta-stream generation are handled by \streamDaQ~ and abstracted from the user, who only writes pure Python. This design enables rapid adaptation to domain-specific requirements, while maintaining the benefits of our streaming quality model.

\subsection{Integration and Usage}
\label{subsec:integration}
\streamDaQ~ provides a straightforward Python interface that allows data practitioners to implement quality monitoring with minimal setup overhead. The framework's design emphasizes simplicity while maintaining flexibility, enabling users to focus on defining quality requirements rather than managing streaming infrastructure.
\autoref{lst:simple-api} presents a self-contained source code example, highlighting the conciseness and simplicity of \streamDaQ's API. Window definition follows an intuitive pattern, while quality checks can be composed by selecting from \streamDaQ's comprehensive suite of built-in functionalities. This Python-native approach enables seamless integration with existing data science workflows. Users can incorporate quality monitoring into their pipelines while leveraging familiar Python syntax and established libraries. For example, quality monitoring can be easily integrated with pandas-based data preprocessing, scikit-learn model training pipelines, or Jupyter notebook-based exploratory analysis. The framework's design eliminates the need for complex distributed system configurations or specialized streaming knowledge, making real-time quality monitoring accessible to the broader data science community already working within the Python ecosystem.

\begin{lstlisting}[style=pythonstyle, caption={Window settings configuration and checks' selection using \streamDaQ's API.}, label={lst:simple-api}]
from streamdaq import DaQMeasures as dqm
from an_existing_workflow import verify_externally

def is_tampere_frequent(loc):
    return True if loc=='Tampere' else False

# tumbling 1h window; 1min cut-off for late data
daq = StreamDaQ().configure(
    window=tumbling(duration=timedelta(hours=1)),
    wait_for_late=timedelta(minutes=1), ...)

# static, custom and external-logic constraints
daq.add(dqm.count('id'), ">10") \
    .add(dqm.most_frequent('location'), 
        lambda loc: is_tampere_frequent(loc)) \
    .add(dqm.min('price'), 
        lambda price: verify_externally(price)) \
    .add(dqm.distinct_count_approx('items'), ...)

# monitoring and access to the meta stream
meta_stream = daq.watch_out()
\end{lstlisting}

\section{Experimental Evaluation}
\label{sec:experiments}
Having demonstrated \streamDaQ's seamless integration into data science workflows, in this section, we evaluate our approach against production-grade alternatives. While our stream-first quality monitoring model introduces novel concepts not directly supported by existing tools, we assess \streamDaQ's practical value through a comprehensive dual evaluation. In particular, we focus on both quantitative performance characteristics (\autoref{subsec:performance_evaluation}) and qualitative functional capabilities (\autoref{subsec:functionality_comparison}) in terms of practical utility for quality monitoring of streaming applications. Below, we outline our evaluation strategy, which addresses the challenges of comparing against approaches that were originally designed to solve a non-identical problem to ours.

\vspace{\gapBetweenPars}\noindent\textbf{Evaluation Strategy.}
Data quality monitoring for unbounded streams has received limited attention in existing literature, despite significant advances in static data quality tools and methodologies. The tools we evaluate against, Amazon's Deequ~\cite{Schelter2018, deequ_github} and Apache Griffin~\cite{griffin_github}, were not originally designed for stream quality monitoring but can potentially be adapted through their underlying frameworks.

We selected Deequ as our primary quantitative comparison baseline for the following reasons. First, it represents a mature, production-grade solution with a comprehensive suite of quality checks for static data. Second, while primarily targeting static and evolving datasets, Deequ can be adapted for streaming scenarios through Spark Streaming integration, making it a relevant performance baseline. Third, its distributed computing approach provides a meaningful baseline for our lightweight, Python-native design. To ensure fair comparison, our evaluation utilizes only a subset of \streamDaQ's model expressiveness that corresponds to functionality available in Deequ's static data capabilities. We detail our approach for adapting Deequ to streaming contexts in \autoref{subsec:performance_evaluation}.

Apache Griffin was excluded from quantitative evaluation due to fundamental architectural differences that prevent meaningful performance comparison. While Griffin supports both batch and streaming modes, its streaming capabilities are severely limited: only one check (accuracy) is documented in its official website~\cite{griffin_streaming_website} as applicable to streaming, requiring reference datasets to validate target streams. This renders such a check impractical for unbounded data scenarios. The profiling check is documented as applicable to streaming in the tool's GitHub repository~\cite{griffin_streaming_measures}, producing single aggregated values persisted when streams terminate. This represents a fundamentally different approach from our continuous window-based monitoring. Adapting Griffin to support window-based continuous assessment would require extensive source code refactoring for only two checks, which is beyond the scope of this evaluation. However, we do provide comprehensive qualitative comparison with Griffin's documented capabilities in \autoref{subsec:functionality_comparison}.

These observations reinforce our motivation for \streamDaQ: the need for a practical, self-contained quality monitoring solution that combines rich functionality with straightforward deployment, accessible for all. Our evaluation strategy thus focuses on demonstrating both the performance advantages of our stream-first model implementation and the practical utility of our model's expressiveness compared to adapted static approaches.

\subsection{Experimental Setup and Results}
\label{subsec:performance_evaluation}

\vspace{\gapBetweenPars}\noindent\textbf{Main Setup Details.}
Experiments were run in isolated Docker containers, each configured with defined resources limits: 8--16 GB for memory and 2--8 cores for CPU utilization. Our host machine runs Ubuntu 24.04 LTS with 12 cores, 32GB RAM, and 500GB SSD storage. Both systems used default batch settings without additional tuning to reflect out-of-the-box performance. For reproducibility, our implementation and experimental code are available on GitHub\footnote{\url{https://github.com/Bilpapster/Stream-DaQ-experiments}}.
\streamDaQ~ experiments were conducted using Python 3.12.5 and Pathway 0.16.0. For Deequ comparisons, we used Apache Spark 3.5.0 with Scala 2.12.10 and Deequ 2.0.8, built using Maven. We adapted Deequ to perform checks on distributed, windowed Spark DataFrames using the Spark Streaming API. This adaptation, albeit not straightforward, is essentially a wrapper for continuously calling Deequ's assessment functions on the (distributed) records of each window, which is handled by Spark Streaming. Our adaptation code is also hosted in our experimental GitHub repository. All experiments used Apache Kafka 2.8 with Zookeeper 3.7.0 as the streaming source to ensure fair comparison, since Spark (and therefore Deequ) requires Kafka for its streaming capabilities. 

We employed the Reddit Comments Dataset\footnote{\url{https://www.kaggle.com/datasets/kaggle/reddit-comments-may-2015}}, a real-world streaming dataset consisting of 22 features, representing live tracking of comments posted on the Reddit platform during May 2015. We selected this dataset since it was also used for evaluation purposes by the authors of Deequ~\cite{Schelter2018}, although in an evolving-data context. In our setting, we replay the comments data stream using the provided creation timestamp as event time for each record. To evaluate scalability, we created finite streams of varying number of input records from 100K to 500K. We also used some smaller streams before as a warm-up mechanism, so that any Spark-related initialization for Deequ's execution is discarded.

\vspace{\gapBetweenPars}\noindent\textbf{Window Configurations and Metrics.}
We conducted experiments using multiple window settings: 1-, 5-, 30-, and 60-minute \textbf{tumbling} windows; and 5-, 10-, and 30-minute \textbf{sliding} windows with slides of 50\% and 10\% for all durations. The average number of records per window is
1783, 2276, 5731, 20035, and 45107 respectively.
For each input stream, we measured overall throughput and latency, as well as per-window performance indicators: average total execution time and net processing time per window. For Deequ, we measure net processing time \emph{excluding Spark's windowing}, while for \streamDaQ~ the net processing time stands for both windowing and execution of quality checks. All numbers reported are average values over 5 or more runs.

\begin{figure*} [tb!]
  \centering
  \includegraphics[width=\linewidth]{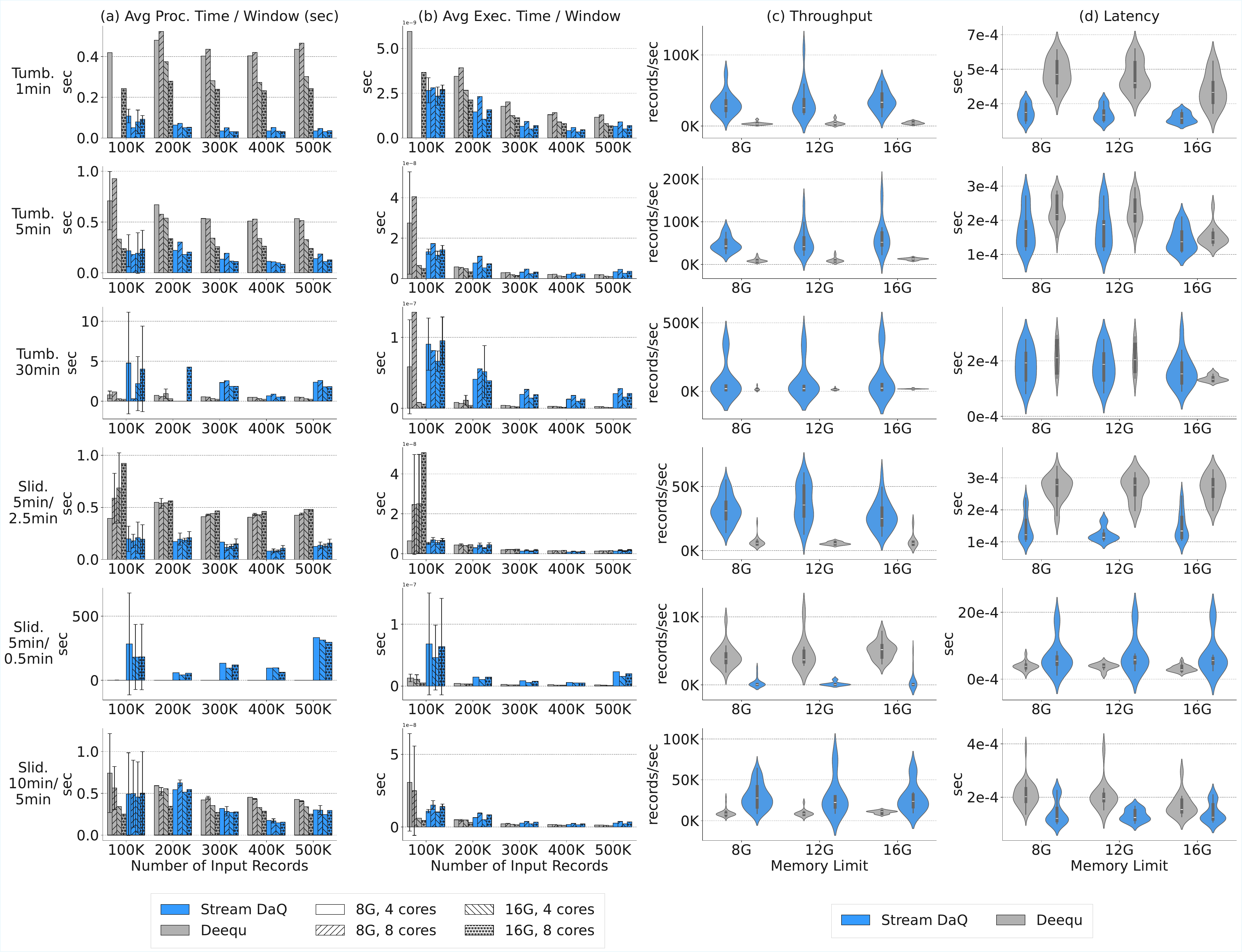}
  \caption{Comparison between \streamDaQ~ and Amazon's Deequ~\cite{Schelter2018} across varying window settings, number of input records and resource constraints on single machine for the Reddit comments dataset.}
  \label{fig:per_window_and_per_CPU_util}
\end{figure*}

\begin{figure} [tb!]
  \centering
  \includegraphics[width=\linewidth]{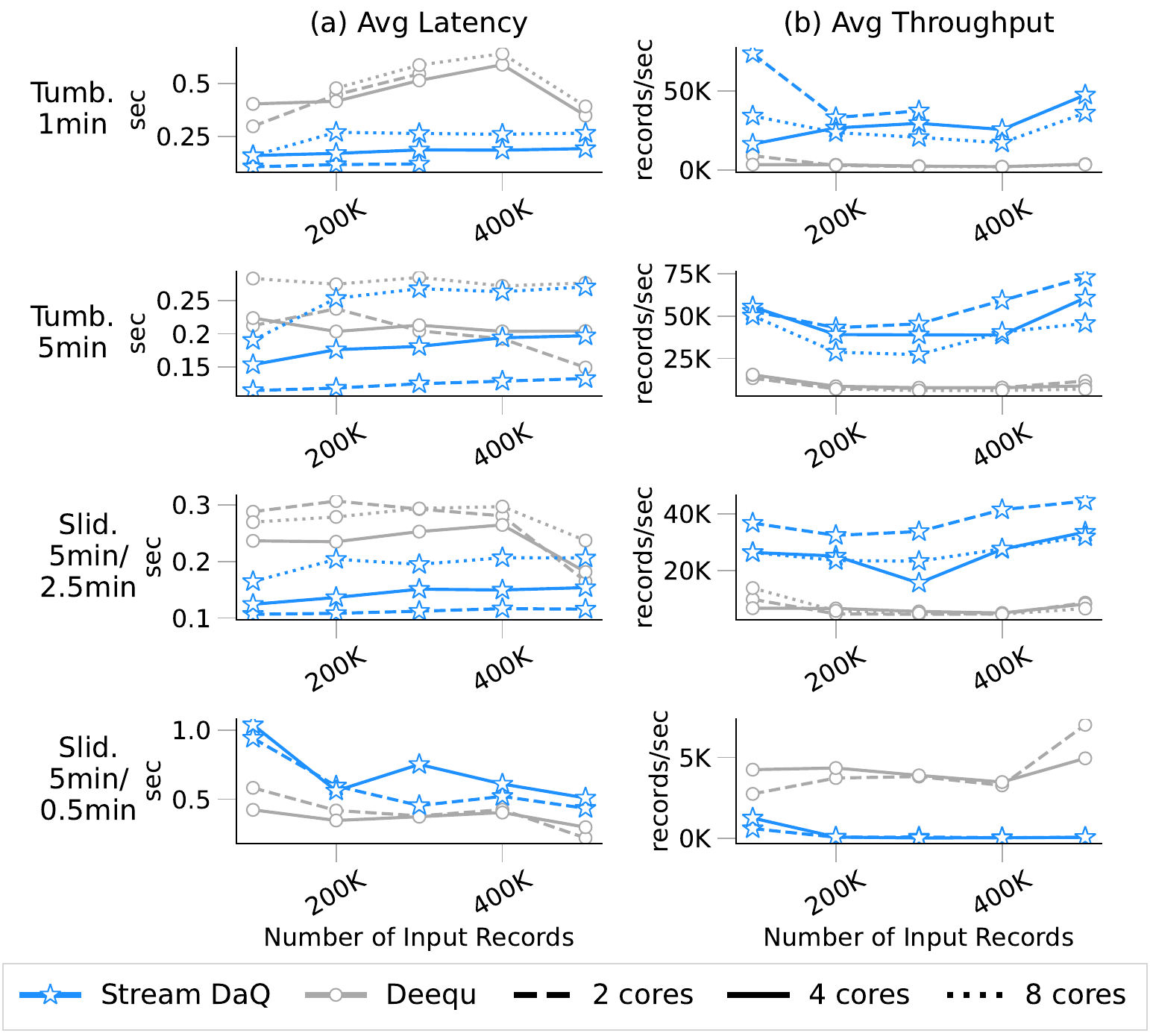}
  \caption{Latency (left) and throughput (right) comparison between \streamDaQ~ and Amazon's Deequ~\cite{Schelter2018} over increasing number of input records from the Reddit comments dataset.}
  \label{fig:latency_throughput}
\end{figure}

\vspace{\gapBetweenPars}\noindent\textbf{Results.}
Our experimental evaluation demonstrates that \streamDaQ~ significantly outperforms Deequ across all configurations that align with real-time streaming quality monitoring requirements. As shown in Figures~\ref{fig:per_window_and_per_CPU_util} and ~\ref{fig:latency_throughput}, \streamDaQ's performance advantages are most pronounced \textbf{for smaller window configurations}, achieving superior throughput and lower latency across all tested stream sizes for 1-minute and 5-minute tumbling windows. Even when Deequ utilizes additional CPU cores or memory resources, \streamDaQ~ maintains its performance advantage with more constrained resources, demonstrating the efficiency of our lightweight, Python-native approach. Observed improvements are up to 13.8 times.
\textbf{For larger window durations} (30-minute and 60-minute tumbling windows) \streamDaQ~ demonstrates comparable performance, maintaining competitive latency and throughput on average. These scenarios, involving processing larger batches at once, are better handled by the batch-oriented, distributed computations of Deequ. However, they do not directly align with real-time quality monitoring requirements, where many small continuous operations are the key to timely, accurate detection of temporally dependent data stream errors. 

Sliding window configurations reveal nuanced performance characteristics that align with different streaming scenarios. \textbf{For 50\% overlap configurations} (simulating moderate temporal granularity), \streamDaQ~ maintains clear performance advantages similar to tumbling windows. However, \textbf{for 90\% overlap} scenarios (requiring extensive recomputation), performance becomes more comparable between systems. This slight performance degradation denotes that there is room of improvement of the incremental nature of \streamDaQ~ processing, but should be also considered against the superior functional expressiveness that \streamDaQ~ offers.

Overall, these results validate our core thesis: \streamDaQ's stream-first architecture provides substantial performance benefits precisely where they matter most—in real-time monitoring scenarios requiring rapid window processing and immediate quality assessment. While maintaining competitive performance across all configurations, \streamDaQ~ uniquely combines efficiency with the comprehensive functionality enabled by our streaming quality monitoring model, making it particularly suitable for continuous data quality assessment in production environments.

\subsection{Functionality Comparison} \label{subsec:functionality_comparison}
Complementary to our performance analysis, we present here an extensive qualitative evaluation, systematically comparing \streamDaQ's capabilities against existing streaming-capable tools to demonstrate the practical advantages of our quality monitoring model.

\vspace{\gapBetweenPars}\noindent\textbf{Comparison with Apache Griffin.} 
Our investigation of Apache Griffin proved challenging due to inconsistent, often contradictory documentation among its 
official website\footnote{\url{https://griffin.apache.org}},
GitHub repository\footnote{\url{https://github.com/apache/griffin}},
and Docker image\footnote{\url{https://github.com/apache/griffin/blob/master/griffin-doc/docker/griffin-docker-guide.md}}.
To address these challenges, we systematically examined all documentation resources comparatively. To eliminate any biased decisions, we assume full implementation of the all sparsely documented features found in at least one resource. Even so, Griffin's capabilities remain severely constrained.

For streaming mode, only two measures are documented: \textbf{accuracy}, limited to exact matching between reference and target streams, and \textbf{profiling}, producing single aggregated values when streams terminate. For batch mode, Griffin additionally supports \textbf{completeness} through custom SQL expressions only, \textbf{duplication}, focused on static data concepts without streaming adaptations, and \textbf{schema conformance} restricted to data type validation. In contrast, our model's check categories (\autoref{subsec:check_categories}) provide a much more comprehensive, practical coverage. Tuple-at-a-time checks enable immediate validation, while window context checks support sophisticated continuous aggregations flowing as a meta-stream. Reference data checks offer flexible baseline comparisons without making impractical assumptions about the existence of a high-quality stream replica. At the same time, \streamDaQ~ accounts for the presence, absence and order of columns beyond type checking, while regarding duplication,  explicitly provides temporally fine-grained detection through both uniqueness and distinctness; built-in checks that can additionally be keyed. This compositional expressiveness (\autoref{subsec:check_composition}) allows combining all these categories to address even more complex scenarios that Griffin's limited measures cannot handle. Finally, Griffin's approach to extensibility --pushing users towards custom SQL expressions for new quality scenarios-- contrasts sharply with \streamDaQ's philosophy. Where Griffin requires users to implement ad-hoc solutions, \streamDaQ~ provides a comprehensive, ready-to-use suite of quality checks that users can selectively apply to their specific contexts. In case highly customized checks are required, users need only provide the logic in pure Python, typically more expressive and familiar than Spark SQL for modern data scientists.

\vspace{\gapBetweenPars}\noindent\textbf{Comparison with Deequ.} 
Our comparison with Deequ reveals fundamental differences at two distinct levels. First, at the check suite level, \streamDaQ~ provides a superset of Deequ's capabilities enriched with extensive configuration options (\autoref{tab:quality_checks}). Beyond the stream-specific checks absent in Deequ, such as stream freshness validation, value ordering detection, cross-interval validation, and consistent placeholder checking, \streamDaQ~ enhances existing check types with meaningful options (e.g., proper/improper subset validation versus Deequ's simple set membership).

More fundamentally, our model's check categories unlock expressiveness that addresses the dynamic nature of data streams. While Deequ supports evolving datasets through append-only updates with total aggregations since stream inception, our dynamic constraint checks enable temporally fine-grained context horizons. This allows monitoring based on recent, relevant information (e.g., last hour's patterns) rather than entire stream history, directly addressing the temporal volatility of streaming data. Finally, \streamDaQ's meta-stream architecture (\autoref{def:meta_stream}) provides continuous quality insights for downstream consumption, contrasting with Deequ's file-based metrics repository approach. This architectural difference reflects our model's stream-first design versus Deequ's batch-extended paradigm, enabling quality-aware data pipelines that adapt in real-time to quality variations.

\section{How to use \streamDaQ}
\label{sec:how_to_use}
\streamDaQ~ is designed for seamless integration into modern data science workflows, offering flexible input/output capabilities that align with real-world requirements. On the input side, \streamDaQ~ can ingest data from various sources including CSV files, JSON lines, and Kafka topics. More importantly, through its custom Python connector (wrapping Pathway's respective implementations), \streamDaQ~ can process any data that can be parsed with Python—from REST API responses to specialized scientific formats, making it particularly valuable for diverse data science scenarios.

The framework's output capabilities are equally versatile. Quality monitoring results can be persisted to the file system for offline analysis, published to Kafka topics for real-time dashboards, or trigger custom Python callbacks for automated actions. This flexibility, combined with \streamDaQ's comprehensive suite of quality checks (see~\autoref{tab:quality_checks} and ~\autoref{lst:simple-api}), enables various integration patterns: data scientists can incorporate quality monitoring into their Jupyter notebooks for interactive analysis, MLOps engineers can automate model retraining based on input data quality, data engineers can integrate quality alerts into their existing monitoring infrastructure, and so on.
Through this input/output flexibility, \streamDaQ~ seamlessly fits into existing data pipelines as a plug-and-play component, while maintaining its lightweight nature; all achievable through a few lines of Python code.

\section{Implementation Challenges and Choices}
\label{sec:why_streaming_python}

In this section we elaborate on the challenges we encountered while implementing \streamDaQ, discussing the choices we made. We attempt to answer the questions \emph{Why Python?} (\autoref{subsec:python_data_science}) and \emph{Why Pathway?} (\autoref{subsec:why_pathway}) for \streamDaQ. We start with an overview of traditional, well-established stream processing frameworks (\autoref{subsec:traditional_stream_processing}) to understand their benefits and limitations.

\subsection{Stream Processing Landscape}
\label{subsec:traditional_stream_processing}
Stream processing systems have evolved significantly over the past three decades~\cite{Fragkoulis2023}, transitioning through distinct generations that reflect changing requirements and technological capabilities. The first generation (1992-2003) emerged from database systems, focusing on continuous queries and window-based operations. The second generation (2004-2019) marked a shift towards distributed, scale-out architectures designed for massive data processing, while the current third generation (2020-) emphasizes cloud integration and edge computing. Apache Flink~\cite{Carbone2015_flink}, Apache Spark Streaming~\cite{Zaharia2013}, and Apache Storm~\cite{Toshniwal2014} represent the mature outcomes of second-generation systems, each built around sophisticated distributed architectures. Flink provides sophisticated windowing and out-of-order processing. Spark Streaming extends batch processing to handle streaming workloads through micro-batching, while Storm emphasizes low-latency processing. While these systems excel at large-scale data processing, their architectural complexity often exceeds the requirements of focused streaming applications~\cite{QaaD2023}, such as quality monitoring or real-time analytics.

The third generation recognizes the need for more accessible solutions~\cite{Fragkoulis2023}, particularly as streaming applications become ubiquitous across different domains~\cite{Golab2010}. Traditional frameworks, built around JVM-based languages and distributed computing principles, create significant barriers through their complex deployment requirements and indirect Python support. This misalignment with modern data science workflows, where Python dominates, has created opportunities for more focused solutions that maintain essential streaming capabilities while prioritizing accessibility, infrastructure costs, and ease of use. Such solutions enable practitioners to concentrate on domain-specific tasks, decoupling efficient stream processing from the traditional distributed frameworks.

\subsection{Python's Role in Modern Data Science}
\label{subsec:python_data_science}
Python has established itself as the dominant programming language in data science and machine learning~\cite{Castro2023}, consistently ranking first in the PYPL~\cite{pypl_index} (since 2019) and the TIOBE~\cite{tiobe_index} (since 2022) indices. This widespread adoption is particularly evident in its significant growth trajectory, with Python showing a remarkable increase of nearly 10\% in popularity during 2024 alone.

The language's success in data science primarily stems from its rich ecosystem of specialized libraries. NumPy for numerical computations, Pandas for data manipulation, scikit-learn for machine learning and a multitude of others; all enable straightforward and rapid implementation of complex analytical workflows~\cite{Castro2023}. More importantly, Python's accessibility makes it particularly suitable for data scientists, who are not necessarily experienced programmers, allowing them to craft a working prototype in a few lines of code, rather than addressing complex programming concepts.

This combination of accessibility and powerful capabilities has created a positive feedback loop: as more data science tools are developed in Python, the ecosystem becomes richer, further consolidating Python's position as the de facto language for data-intensive tasks. However, when moving to scale, particularly in streaming contexts, data scientists often face significant barriers when trying to use traditional streaming frameworks that are primarily designed around JVM-based languages.
This gap between Python's dominance in data science and the JVM-centric nature of mature streaming frameworks creates an opportunity for frameworks that can bridge these worlds: efficient stream processing in pure Python. 

\subsection{Why Pathway for \streamDaQ?}
\label{subsec:why_pathway}
The growing need for Python-native streaming solutions has led to the emergence of several frameworks that aim to bridge the gap between traditional JVM-based streaming systems and Python-centric data science workflows. Our thorough investigation of these frameworks, summarized in~\autoref{tab:framework_comparison}, reveals distinct approaches in terms of implementation, window support, and deployment requirements.

\vspace{\gapBetweenPars}\noindent\textbf{Pure Python implementations.}
While offering native integration, they often face fundamental limitations. Faust~\cite{faust_github} represents a particularly challenging case: the original project, initiated by Robinhood, is now inactive, with the community maintaining a mirror version. While it supports basic windowing operations (tumbling and sliding windows), its tight coupling with Kafka Streams requires Kafka deployment for any streaming application. Quix~\cite{quix_github}, though actively maintained, shares similar constraints: its windowing operations are restricted to event time retrieved from Kafka topics. Moreover, its pure Python backend suffers from the Global Interpreter Lock (GIL)~\cite{Castro2023}, leading to inherent performance limitations (e.g., low instruction-level parallelism~\cite{Ismail_2018}).

\vspace{\gapBetweenPars}\noindent\textbf{Hybrid implementations.}
Recent frameworks have adopted high-performance backends while maintaining Python accessibility, with a view to combine the best of both worlds. Bytewax~\cite{bytewax_github}, like our chosen framework Pathway, implements its core engine in Rust while exposing a Python API. It supports all three window types (tumbling, sliding, and session) and avoids Kafka dependencies. However, its workflow-oriented API introduces complexity in window operations, particularly in accessing window metadata such as start and end timestamps. Pathway~\cite{pathway2023, pathway_github} distinguishes itself through several key advantages. Its Rust backend, based on Differential Dataflow~\cite{akidau_2015_dataflow}, ensures efficient execution while maintaining a pure Python API for accessibility. It provides comprehensive window support with straightforward metadata access and includes built-in reducers (implemented in Rust) for common operations like count, max, or min —features particularly valuable for data quality measurements. Also, as an actively maintained open-source project, Pathway continues to evolve.

\begin{table}[tb]
    \tablesStretch
    \caption{Python Stream Processing Framework Comparison}
    \label{tab:framework_comparison}
    \resizebox{\linewidth}{!}{%
        \begin{tabular}{lllccccccc}
            \toprule
            \multirow[c]{3}{*}{\textbf{Framework}} & \multirow[c]{3}{*}{\textbf{Version}} & \multirow[c]{3}{*}{\textbf{Backend}} & \multicolumn{4}{c}{\textbf{Window Support}} & \multicolumn{2}{c}{\textbf{Pros}}          & \multicolumn{1}{c}{\textbf{Depend.}}                                                                                                                                                                                               \\
                                                   &                                      &                                      & \multicolumn{4}{c}{\downbracefill}          & \multicolumn{2}{c}{\downbracefill}         & \downbracefill                                                                                                                                                                                                                     \\
                                                   &                                      &                                      & \rotatebox{\rotateHeaders}{\textbf{Tumbl.}} & \rotatebox{\rotateHeaders}{\textbf{Slid.}} & \rotatebox{\rotateHeaders}{\textbf{Sess.}} & \rotatebox{\rotateHeaders}{\textbf{Count}} & \rotatebox{\rotateHeaders}{\textbf{Reducers}} & \rotatebox{\rotateHeaders}{\textbf{Active}} & \rotatebox{\rotateHeaders}{\textbf{Kafka}} \\
            \midrule
            Faust~\hfill\cite{faust_github}        & 0.11.3                               & Python                               & \cmark                                      & \cmark                                     & \xmark                                     & \xmark                                     & \xmark                                        & \xmark                                      & \cmark                                     \\
            Quix~\hfill\cite{quix_github}          & 3.6.0                                & Python                               & \cmark                                      & \cmark                                     & \xmark                                     & \xmark                                     & \xmark                                        & \cmark                                      & \cmark                                     \\
            Bytewax~\hfill\cite{bytewax_github}    & 0.21.1                               & Rust                                 & \cmark                                      & \cmark                                     & \cmark                                     & \xmark                                     & \xmark                                        & \cmark                                      & \xmark                                     \\
            Pathway~\hfill\cite{pathway2023}       & 0.16.4                               & Rust                                 & \cmark                                      & \cmark                                     & \cmark                                     & \xmark                                     & \cmark                                        & \cmark                                      & \xmark                                     \\
            \bottomrule
        \end{tabular}
    }
\end{table}

\section{Related Work}
\label{sec:background_related_work}

\vspace{\gapBetweenPars}\noindent\textbf{Tools for Static Data.}
Several tools have been developed for assessing static, relational data. The Data Quality Assessment Framework~\cite{Coleman_2013} provides structured quality measurement through a rich set of metrics and assessment methods for batch processing. DQA~\cite{Shrivastava2019} offers automated, interactive data quality assessment with executable validation graphs supporting both generic checks (e.g., missing values) and domain-specific validations (e.g., frequency-related checks on time series) through customizable rules. Building upon DQA's validators, DQDF~\cite{Sinthong2022} introduces a data-quality-aware extension to Python Pandas~\cite{pandas_documentation} dataframes, optimizing quality assessment for evolving datasets by detecting metadata changes and avoiding unnecessary computations. CheDDaR~\cite{Restat2023} provides flexible quality evaluation metrics during preprocessing to assist domain experts in choosing appropriate checks. TsQuality~\cite{Qiu2023}, while targeting time-series data in Apache IoTDB~\cite{iot_db}, operates in batch mode on stored data points. These tools' shared limitation lies in their design for unchanged data rather than streaming contexts.

\vspace{\gapBetweenPars}\noindent\textbf{Tools with Streaming Capabilities.}
Some tools primarily designed for static data have attempted to extend their capabilities to streams by leveraging their underlying computing frameworks. However, these approaches fundamentally treat streaming as a batch processing extension, rather than providing native support for window-based monitoring at user-driven temporal granularity.

Amazon's Deequ~\cite{Schelter2018}, built on Apache Spark~\cite{Zaharia2012spark}, provides differential computation for evolving datasets through algebraic states~\cite{Schelter2019}, but its streaming support is limited to Spark's micro-batch model without explicit windowing for continuous assessment. Similarly, SparkDQ~\cite{Gu2021} leverages Spark's distributed processing with primary focus on batch scenarios. Apache Griffin, built on Hadoop and Spark, conceptually supports streaming but offers only two impractical streaming checks~\cite{griffin_streaming_measures}. Luzzu~\cite{luzzu_2016} attempts to address Open Linked Data quality assessment focusing on static and evolving public datasets. A ``streaming'' implementation is also provided, however this is more of a (Java) iterator over finite RDF triples for scalability; fundamentally different from a stream-first, window-based approach for continuous monitoring of unbounded streams, like the one we describe in this paper.

These existing approaches, while offering some streaming capabilities, essentially treat streaming as an extension of batch processing. They apply quality checks on micro-batches without inherent conceptual support for continuous monitoring through windowing. Their reliance on heavyweight frameworks like Spark and Hadoop can introduce significant overhead, proving inefficient for workflows comprising many small, continuous operations~\cite{QaaD2023}; typical in quality monitoring scenarios~\cite{Gassman1995}. Moreover, these frameworks often represent a costly overkill for many real-world applications~\cite{linthicum2024cloud}, requiring complex infrastructure setups and expertise in distributed systems. This creates a significant barrier to entry for data scientists who need straightforward quality monitoring solutions. In contrast, \streamDaQ~ is designed from the ground up for real-time streaming scenarios, offering native support for continuous quality assessment through explicit window-based operations on streams, all within the familiar and widely-adopted Python ecosystem.

\section{Conclusion and Future Work}
\label{sec:conclusion}
This paper presents a novel data quality monitoring model specifically designed for unbounded data streams, addressing fundamental limitations of existing static and incremental approaches. Our model introduces stream-first concepts including configurable windowing mechanisms, dynamic constraint adaptation, and continuous assessment that produces quality meta-streams for real-time pipeline awareness. To demonstrate practical applicability, we developed \streamDaQ, an open-source Python framework that implements our theoretical model while providing seamless integration into existing data science workflows. Our evaluation demonstrates both significant performance advantages over existing production-grade solutions (e.g., 13.8 faster processing of 1 min windows) and richer, stream-first quality checks.

Regarding future work, apart from enhanced incremental processing, further enriching \streamDaQ's check suite and customization options could be highly beneficial to the broader data science community. Taking it a step further, we aim to systematically identify domain-specific quality checks by quantifying their impact on downstream tasks. This approach will help to establish targeted quality monitoring strategies for specific scenarios, such as advanced RAG pipelines, where quality requirements can vary significantly based on the context.

\enlargethispage{4mm}
\bibliographystyle{ACM-Reference-Format}
\bibliography{main}

\end{document}